\documentclass[twocolumn]{aastex631}
\usepackage{hyperref}
\usepackage{graphicx}
\usepackage{natbib}

\begin{document}

\title{JWST-TST DREAMS: Quartz Clouds in the Atmosphere of WASP-17b}


\author[0000-0001-5878-618X]{David Grant}
\affiliation{University of Bristol, HH Wills Physics Laboratory, Tyndall Avenue, Bristol, UK}

\author[0000-0002-8507-1304]{Nikole K. Lewis}
\affiliation{Department of Astronomy and Carl Sagan Institute, Cornell University, 122 Sciences Drive, Ithaca, NY 14853, USA}

\author[0000-0003-4328-3867]{Hannah R. Wakeford}
\affiliation{University of Bristol, HH Wills Physics Laboratory, Tyndall Avenue, Bristol, UK}

\author[0000-0003-1240-6844]{Natasha E. Batalha}
\affiliation{NASA Ames Research Center, Moffett Field, CA, 94035, USA}

\author[0000-0002-5322-2315]{Ana Glidden}
\affiliation{Department of Earth, Atmospheric and Planetary Sciences, Massachusetts Institute of Technology, Cambridge, MA 02139, USA}
\affiliation{Kavli Institute for Astrophysics and Space Research, Massachusetts Institute of Technology, Cambridge, MA 02139, USA}

\author[0000-0002-8515-7204]{Jayesh Goyal}
\affiliation{School of Earth and Planetary Sciences (SEPS), National Institute of Science Education and Research (NISER), HBNI, Odisha, India}

\author[0000-0003-0814-7923]{Elijah Mullens}
\affiliation{Department of Astronomy and Carl Sagan Institute, Cornell University, 122 Sciences Drive, Ithaca, NY 14853, USA}

\author[0000-0003-4816-3469]{Ryan J. MacDonald}
\affiliation{Department of Astronomy, University of Michigan, 1085 S. University Ave., Ann Arbor, MI 48109, USA}

\author[0000-0002-2739-1465]{Erin M. May}
\affiliation{Johns Hopkins APL, 11100 Johns Hopkins Rd, Laurel, MD 20723, USA}

\author[0000-0002-6892-6948]{Sara Seager}
\affiliation{Department of Earth, Atmospheric and Planetary Sciences, Massachusetts Institute of Technology, Cambridge, MA 02139, USA}
\affiliation{Kavli Institute for Astrophysics and Space Research, Massachusetts Institute of Technology, Cambridge, MA 02139, USA} 
\affiliation{Department of Aeronautics and Astronautics, MIT, 77 Massachusetts Avenue, Cambridge, MA 02139, USA}

\author[0000-0002-7352-7941]{Kevin B. Stevenson}
\affiliation{Johns Hopkins APL, 11100 Johns Hopkins Rd, Laurel, MD 20723, USA}

\author[0000-0003-3305-6281]{Jeff A. Valenti}
\affiliation{Space Telescope Science Institute, 3700 San Martin Drive, Baltimore, MD 21218, USA}

\author[0000-0001-6627-6067]{Channon Visscher}
\affiliation{Chemistry \& Planetary Sciences, Dordt University, Sioux Center, IA, USA}
\affiliation{Center for Extrasolar Planetary Systems, Space Science Institute, Boulder, CO, USA}

\author[0000-0001-8703-7751]{Lili Alderson}
\affiliation{University of Bristol, HH Wills Physics Laboratory, Tyndall Avenue, Bristol, UK}

\author[0000-0002-0832-710X]{Natalie H. Allen}
\affiliation{William H. Miller III Department of Physics and Astronomy, Johns Hopkins University, Baltimore, MD 21218, USA}

\author[0000-0003-4835-0619]{Caleb I. Ca\~{n}as}
\affiliation{NASA Goddard Space Flight Center, Greenbelt, MD 20771, USA}

\author[0000-0001-8020-7121]{Knicole Col\'{o}n}
\affiliation{NASA Goddard Space Flight Center, Greenbelt, MD 20771, USA}

\author[0000-0003-4003-8348]{Mark Clampin}
\affiliation{NASA Headquarters, 300 E Street SW, Washington, DC 20546, USA}

\author[0000-0001-9513-1449]{N\'{e}stor Espinoza}
\affiliation{Space Telescope Science Institute, 3700 San Martin Drive, Baltimore, MD 21218, USA}
\affiliation{William H. Miller III Department of Physics and Astronomy, Johns Hopkins University, Baltimore, MD 21218, USA}

\author[0000-0003-0854-3002]{Am\'{e}lie Gressier}
\affiliation{Space Telescope Science Institute, 3700 San Martin Drive, Baltimore, MD 21218, USA}

\author[0000-0001-5732-8531]{Jingcheng Huang}
\affiliation{Department of Earth, Atmospheric and Planetary Sciences, Massachusetts Institute of Technology, Cambridge, MA 02139, USA}

\author[0000-0003-0525-9647]{Zifan Lin}
\affiliation{Department of Earth, Atmospheric and Planetary Sciences, Massachusetts Institute of Technology, Cambridge, MA 02139, USA}

\author[0000-0002-2508-9211]{Douglas Long}
\affiliation{Space Telescope Science Institute, 3700 San Martin Drive, Baltimore, MD 21218, USA}

\author[0000-0002-2457-272X]{Dana R. Louie}
\affiliation{NASA Goddard Space Flight Center, Greenbelt, MD 20771, USA}

\author[0000-0003-2314-3453]{Maria Pe\~{n}a-Guerrero}
\affiliation{Space Telescope Science Institute, 3700 San Martin Drive, Baltimore, MD 21218, USA}

\author[0000-0002-5147-9053]{Sukrit Ranjan}
\affiliation{Lunar and Planetary Laboratory/Department of Planetary Sciences, University of Arizona, Tucson, AZ 85721, USA}

\author[0000-0001-7393-2368]{Kristin S. Sotzen}
\affiliation{Johns Hopkins APL, 11100 Johns Hopkins Rd, Laurel, MD 20723, USA}
\affiliation{William H. Miller III Department of Physics and Astronomy, Johns Hopkins University, Baltimore, MD 21218, USA}

\author[0000-0002-2643-6836]{Daniel Valentine}
\affiliation{University of Bristol, HH Wills Physics Laboratory, Tyndall Avenue, Bristol, UK}

\author[0000-0003-2861-3995]{Jay Anderson}
\affiliation{Space Telescope Science Institute, 3700 San Martin Drive, Baltimore, MD 21218, USA}

\author[0000-0001-6396-8439]{William O. Balmer}
\affiliation{William H. Miller III Department of Physics and Astronomy, Johns Hopkins University, Baltimore, MD 21218, USA}
\affiliation{Space Telescope Science Institute, 3700 San Martin Drive, Baltimore, MD 21218, USA}

\author[0000-0003-3858-637X]{Andrea Bellini}
\affiliation{Space Telescope Science Institute, 3700 San Martin Drive, Baltimore, MD 21218, USA}

\author[0000-0002-9803-8255]{Kielan K. W. Hoch}
\affiliation{Center for Astrophysics and Space Sciences, University of California, San Diego, La Jolla, CA 92093, USA}

\author[0000-0003-2769-0438]{Jens Kammerer}
\affiliation{Space Telescope Science Institute, 3700 San Martin Drive, Baltimore, MD 21218, USA}

\author[0000-0001-9673-7397]{Mattia Libralato}
\affiliation{AURA for the European Space Agency (ESA), Space Telescope Science Institute, 3700 San Martin Drive, Baltimore, MD 21218, USA}

\author{C. Matt Mountain}
\affiliation{Association of Universities for Research in Astronomy, 1331 Pennsylvania Avenue NW Suite 1475, Washington, DC 20004, USA}

\author[0000-0002-3191-8151]{Marshall D. Perrin}
\affiliation{Space Telescope Science Institute, 3700 San Martin Drive, Baltimore, MD 21218, USA}

\author{Laurent Pueyo}
\affiliation{Space Telescope Science Institute, 3700 San Martin Drive, Baltimore, MD 21218, USA}

\author[0000-0003-4203-9715]{Emily Rickman}
\affiliation{European Space Agency (ESA), ESA Office, Space Telescope Science Institute, Baltimore, MD 21218, USA}

\author[0000-0002-4388-6417]{Isabel Rebollido}
\affiliation{Centro de Astrobiolog\'{i}a (CAB, CSIC-INTA), ESAC Campus Camino Bajo del Castillo, s/n, Villanueva de la Ca\~{n}ada, E-28692 Madrid, Spain}

\author[0000-0001-8368-0221]{Sangmo Tony Sohn}
\affiliation{Space Telescope Science Institute, 3700 San Martin Drive, Baltimore, MD 21218, USA}

\author[0000-0001-7827-7825]{Roeland P. van der Marel}
\affiliation{Space Telescope Science Institute, 3700 San Martin Drive, Baltimore, MD 21218, USA}
\affiliation{William H. Miller III Department of Physics and Astronomy, Johns Hopkins University, Baltimore, MD 21218, USA}

\author[0000-0002-1343-134X]{Laura L. Watkins}
\affiliation{AURA for the European Space Agency (ESA), Space Telescope Science Institute, 3700 San Martin Drive, Baltimore, MD 21218, USA}

\begin{abstract}
\noindent Clouds are prevalent in many of the exoplanet atmospheres that have been observed to date. For transiting exoplanets, we know if clouds are present because they mute spectral features and cause wavelength-dependent scattering. While the exact composition of these clouds is largely unknown, this information is vital to understanding the chemistry and energy budget of planetary atmospheres. In this work, we observe one transit of the hot Jupiter WASP-17b with JWST's MIRI LRS and generate a transmission spectrum from 5--12\,\textmu m. These wavelengths allow us to probe absorption due to the vibrational modes of various predicted cloud species. Our transmission spectrum shows additional opacity centered at 8.6\,\textmu m, and detailed atmospheric modeling and retrievals identify this feature as SiO$_2$(s) (quartz) clouds. The SiO$_2$(s) clouds model is preferred at 3.5--4.2$\sigma$ versus a cloud-free model and at 2.6$\sigma$ versus a generic aerosol prescription. We find the SiO$_2$(s) clouds are comprised of small ${\sim}0.01$\,\textmu m particles, which extend to high altitudes in the atmosphere. The atmosphere also shows a depletion of H$_2$O, a finding consistent with the formation of high-temperature aerosols from oxygen-rich species. This work is part of a series of studies by our JWST Telescope Scientist Team (JWST-TST), in which we will use Guaranteed Time Observations to perform Deep Reconnaissance of Exoplanet Atmospheres through Multi-instrument Spectroscopy (DREAMS).
\newline
\end{abstract}

\section{Introduction} \label{sec:intro}
Aerosols -- clouds generated via condensation or hazes via photochemistry -- are fundamental components of planetary atmospheres. Their ubiquity in the Solar System planets has now been extended to encompass many exoplanets \citep{gao2021aerosols}, with indications of their presence from the very first transmission measurements of an exoplanet atmosphere \citep{charbonneau2002detection, fortney2003indirect}. Aerosols may contribute significant opacity, impacting how light is reflected, absorbed, and re-radiated, and characterizing them is therefore crucial to understanding the entire energy budget, chemistry, and dynamics of exoplanet atmospheres.

Observationally, the presence of aerosols is generally inferred by the muting, or entire obscuration, of spectral features \citep[e.g.,][]{kreidberg2015detection, sing2016continuum, wakeford2019muting}. This process can be associated with large ($>$1\,\textmu m-sized) particles which uniformly affect all wavelengths, or by small ($<$1\,\textmu m-sized) particles which preferentially scatter shorter wavelengths in the UV-optical. Whilst there is plentiful evidence for the existence and impact of aerosols, as yet no cloud species have been definitively identified in a transiting exoplanet atmosphere.

\citet{Wakeford2015} showed that uniform opacity or scattering can be linked to the particle size and composition of the clouds, where small sub-micron particles can show prominent absorption at mid-infrared wavelengths. The absorbing wavelengths (typically $>$3\,\textmu m) depend on the major diatomic bond, and its associated vibrational mode, and thereby enable specific cloud species to be identified. \citet{gao2020aerosol} has shown that there may be a correlation between the composition of the aerosol and temperature of a planet’s atmosphere, where below 950\,K aerosols are dominated by hazes, and above this temperature silicate clouds dominate with minor contributions from other oxygen-rich species. Silicates (as individual species or as a mixture of different silica-based grains) have previously been identified in brown-dwarf atmospheres using \textit{Spitzer} spectroscopy \citep[e.g.,][]{cushing2006,saurez2022} and are largely held responsible for the L-T transition which spans typical hot Jupiter temperatures. Identifying these cloud species in exoplanet atmospheres is now possible with JWST’s Mid Infrared Instrument \citep[MIRI,][]{wright2023mid}, the only instrument suitable for transit spectroscopy that provides wavelength coverage beyond 5\,\textmu m, and has already detected the presence of silicates in the directly imaged spectrum of the planetary-mass companion VHS 1256-1257b \citep{miles2023jwst}.

In this study we observed WASP-17b, a hot Jupiter in a retrograde orbit around an F6 star \citep{anderson2010, triaud2010spin} with an orbital period of 3.735 days \citep{Alderson2022}. WASP-17b has a mass, radius, and equilibrium temperature of $0.477 \, \rm{M}_{\rm{Jup}}$, $1.932 \, \rm{R}_{\rm{Jup}}$, and $1771 \, \rm{K}$, respectively \citep{anderson2011, southworth2012homogeneous}. These parameters mean that WASP-17b has a huge atmospheric scale height of ${\sim}2,000 \, \rm{km}$, making it an ideal target for transmission spectroscopy. Furthermore, WASP-17b's tidally-locked orbit and permanent dayside irradiation are predicted to create large day-to-night and morning-to-evening differences in the atmospheric properties \citep{kataria2016atmospheric, zamyatina2023observability}. In particular, temperatures of ${\sim}1300$\,K, at 0.1 mbar pressure levels on the limb, may provide the necessary conditions for the condensation of various cloud species which are detectable with MIRI.

WASP-17b has been previously observed with ground-based high-resolution instruments \citep{wood2011transmission, zhou2012detection, bento2014optical, sedaghati2016potassium, khalafinejad2018atmosphere}, observed with the {\it Spitzer} and Hubble Space Telescopes \citep{mandell2013exoplanet, sing2016continuum, Alderson2022}, and has been included in numerous modeling efforts \citep{barstow2016consistent, fisher2018retrieval, pinhas2019h2o, welbanks2019mass, min2020arcis, Alderson2022, saba2022transmission}. These efforts have yielded strong detections of H$_2$O, a tentative detection of CO$_2$, and a variety of detections and non-detections of Na I and K I. The comprehensive analysis by \citet{Alderson2022}, utilizing all of the available space-based data, presented two results pertinent to this analysis. First, the need to include a uniform/gray cloud deck and wavelength-dependent scattering aerosol prescription to model the data; and second, a bimodality in the retrieved H$_2$O and metallicity values.

This paper is part of a series to be presented by the JWST Telescope Scientist Team (JWST-TST)\footnote{\label{fn-tst-team} \url{https://www.stsci.edu/~marel/jwsttelsciteam.html}}, led by M. Mountain and convened in 2002 following a competitive NASA selection process. In addition to providing scientific support for observatory development through launch and commissioning, the team was awarded 210 hours of Guaranteed Time Observer (GTO) time. This time is being used for studies in three different subject areas: (a) Transiting Exoplanet Spectroscopy (lead: N. Lewis); (b) Exoplanet and Debris Disk High-Contrast Imaging (lead: M. Perrin); and (c) Local Group Proper Motion Science (lead: R. van der Marel). A common theme of these investigations is the desire to pursue and demonstrate science for the astronomical community at the limits of what is made possible by the exquisite optics and stability of JWST. The present paper is part of our work on Transiting Exoplanet Spectroscopy, which focuses on Deep Reconnaissance of Exoplanet Atmospheres using Multi-instrument Spectroscopy (DREAMS) of three transiting exoplanets representative of key classes: Hot Jupiters (WASP-17b, GTO~1353), Warm Neptunes (HAT-P-26b, GTO~1312), and Temperate Terrestrials (TRAPPIST-1e, GTO~1331). Here we present our observational analysis and interpretation of WASP-17b's mid-infrared (5--12\,\textmu m) transmission spectrum.

\section{Observations, Data Reduction, and Spectral Generation} \label{sec:obs-red}
We observed one transit of WASP-17b with JWST's MIRI Low Resolution Spectroscopy (LRS) Slitless mode \citep{kendrew2015} as part of GTO-1353 (PI Lewis) on 12 to 13 March 2023 (program observation 5). For this transit observation we used the MIRI FASTR1 readout pattern and obtained a total of 1,276 integrations, each consisting of 175 groups, for a total exposure time of 35,716~s (9.92~hrs). This observational strategy gave us a time series spanning twice the ${\sim}$4.4-hour transit duration \citep{anderson2010, anderson2011} as well as an additional hour to account for the observation start window and detector settling timescales.

To reduce the WASP-17b JWST MIRI LRS transit observation, we employ two independent pipelines designed for JWST exoplanet time-series observations: ExoTiC-MIRI \citep{grant_david_2023_8211207} and Eureka! \citep{Bell2022}. In the following sections we describe specifics of the reductions performed to generate our JWST MIRI LRS transmission spectra of WASP-17b. We note that the reductions were initially performed independently with limited communication between individuals reducing the observations to ensure the robustness of our results. 

\subsection{Data Reduction: \textsc{ExoTiC-MIRI}}
\label{sec:obs-red-exotic}
The ExoTiC-MIRI\footnote{\url{https://exotic-miri.readthedocs.io/}} pipeline is interoperable with the JWST Science Calibration Pipeline \citep[][\textsc{jwst}]{bushouse_howard_2022_7325378} enabling default processing steps to be switched out, or interleaved, with custom steps when warranted. Starting from the {\it uncal.fits} files we processed the data through the steps \textsc{dq\_init}, \textsc{saturation}, \textsc{custom\_drop\_groups}, \textsc{custom\_linearity}, \textsc{dark\_current}, \textsc{jump}, and \textsc{ramp\_fit}, where the \textsc{custom\_} prefix denotes an ExoTiC-MIRI step. For the default \textsc{jwst} steps, standard settings from pipeline v1.8.2 were used, and CRDS context 1077, except for the jump step where the rejection threshold was increased to 15 to prevent spurious flagging of cosmic rays, and the gain value was set to 3.1 electrons/DN \citep{bell2023first}.

As for the custom steps, the 175 groups are a sufficiently large number such that we could refine the group-level ramps prior to ramp fitting. Inspection of these ramps revealed evidence of the reset switch charge decay (RSCD) effect impacting the first ${\sim}$12 groups, the last-frame effect pulling down the final group \citep{ressler2015mid, wright2023mid}, and residual non-linearity remaining in the groups between. This non-linearity persisted even after the default linearity correction was applied, most likely due to the interplay between the non-linearity effect and debiasing-induced brighter-fatter effect (BFE) changing the apparent sensitivity of each pixel \citep{argyriou2023brighter}. As such, we performed a self-calibration of the linearity correction. The \textsc{custom\_linearity} step utilizes groups 12 to 40 as the presumed linear portion, and then extrapolates a linear fit from this region to derive a correction factor based on deviations of groups 12 to 174. The correction factor takes the form of a fourth-order polynomial where the zeroth- and first-order terms are held fixed at zero and one, respectively. The correction was derived per detector amplifier, essentially segmenting the correction into columns of differing fluence across the core of the point spread function owing to the dependence of the RSCD and BFE on said fluence. The first 12 and final group of each integration were dropped altogether. 

After ramp fitting, the newly created {\it rateimages.fits} files were processed with the steps \textsc{assign\_wcs}, \textsc{src\_type}, \textsc{flat\_field}, \textsc{custom\_clean\_outliers}, \textsc{custom\_background\_subtract}, and \textsc{custom\_extract\_1d}. The custom cleaning step aims to remove any unidentified outliers as well as replace known bad pixels from the data quality arrays. To accomplish this, a spatial profile was estimated from polynomial fits to the detector columns as per optimal extraction \citep{horne1986optimal}, and pixels were iteratively replaced by this profile value if they were more than four standard deviations discrepant or had the data quality flag do\_not\_use. The background step subtracts a row-by-row background using the median value from columns 12 to 22 and 50 to 68. A time series of stellar spectra were then extracted using a fixed-width box aperture. This aperture was centered on column 36 and extended 3 pixels in either direction.

\subsection{Data Reduction: \textsc{Eureka!}}
\label{sec:obs-red-eureka}

\begin{figure*}
\centering
\includegraphics[width=1.0\textwidth]{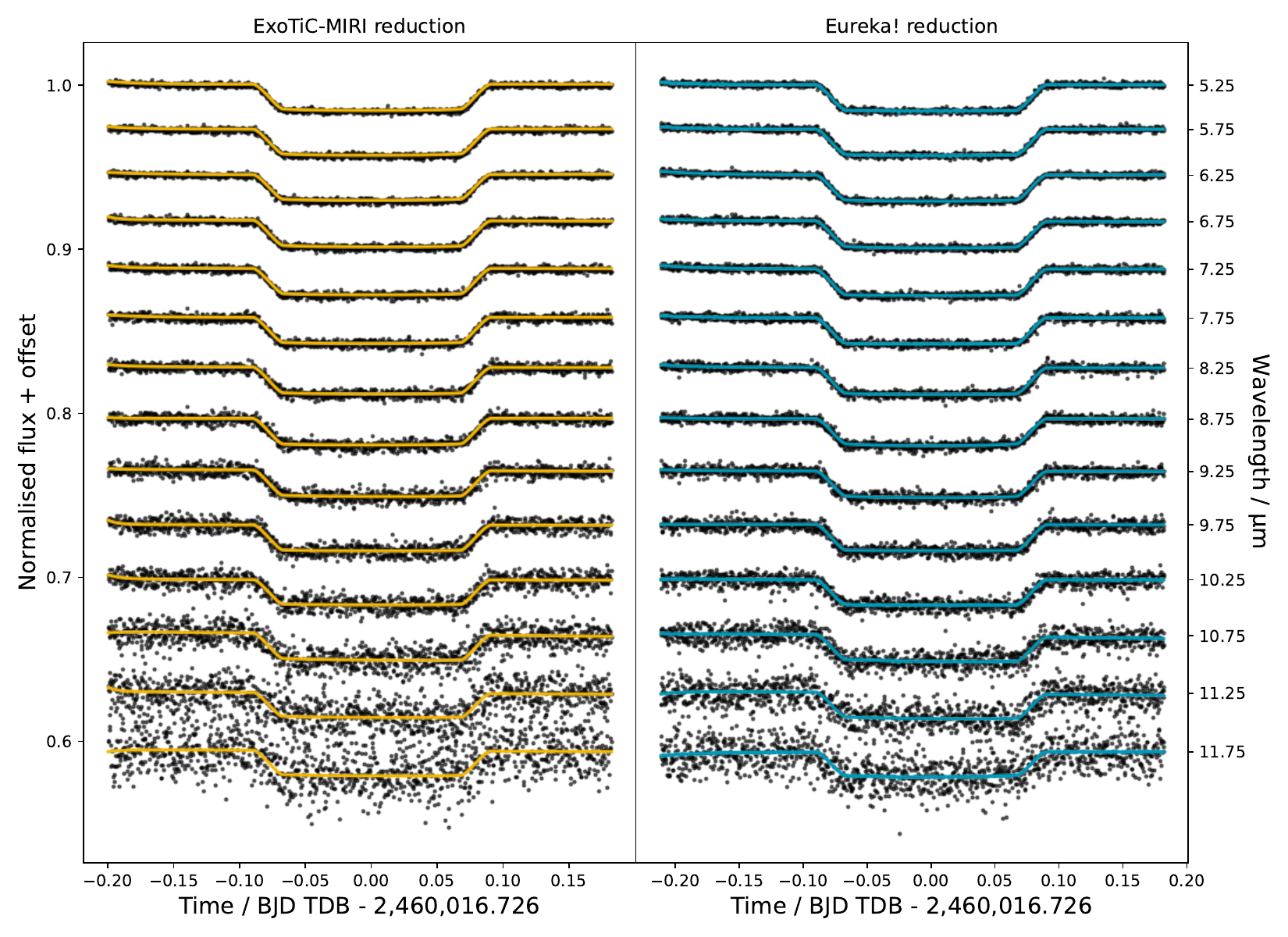}
\caption{ExoTiC-MIRI (left) and Eureka! (right) spectroscopic transit light curves, binned to a resolution of 0.5 \textmu m, and best-fit models. The times on the x-axis have had the centre-of-transit time subtracted and the wavelengths of each light curve are shown on the right-hand y-axis.}
\label{fig:light_curve_waterfall}
\end{figure*}

\begin{figure*}
\includegraphics[width=0.94\textwidth]{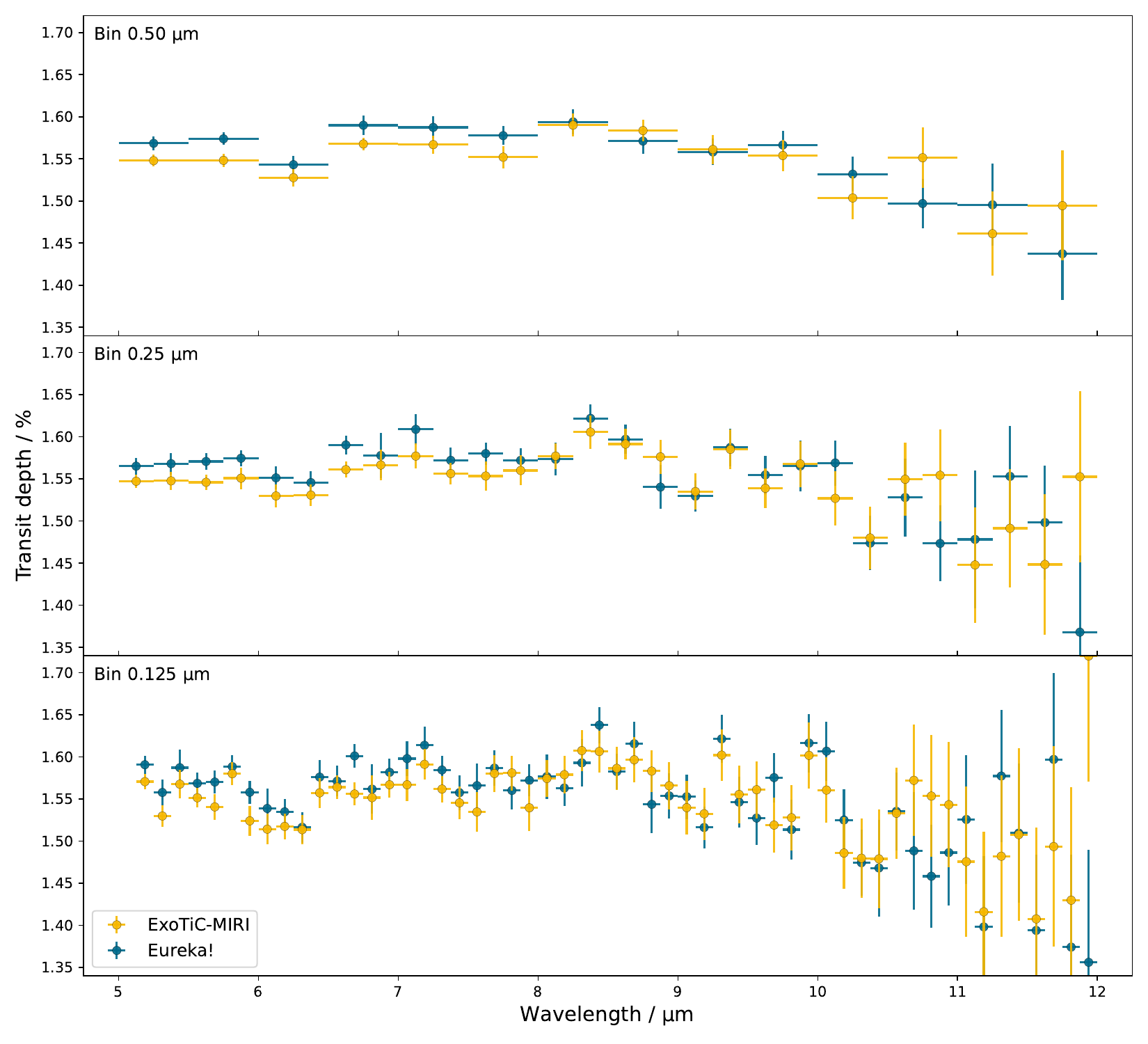}
\caption{Comparison of transmission spectra generated at three resolutions for both the ExoTiC-MIRI (yellow) and Eureka! (blue) reductions. The resolutions correspond to bin widths of 0.50, 0.25, and 0.125 \textmu m. The systematic differences at wavelengths $< 8$\,\textmu m are due to different linearity corrections used in each reduction (see Section \ref{sec:reduction_intercomparison} for details). All data products and models are available at \url{https://doi.org/10.5281/zenodo.8360121}.}
\label{fig:transmission_spectra}
\end{figure*}

The Eureka!\footnote{\url{https://eurekadocs.readthedocs.io/}} reduction pipeline leverages the JWST Science Calibration Pipeline \citep[v1.8.2][\textsc{jwst}]{bushouse_howard_2022_7325378} for stages 1 and 2. We started our Eureka! reduction with the {\it uncal.fits} files. In the stage 1 ``ramps-to-slopes" step, the default JWST MIRI time-series observations settings are applied with the first and last groups of each integration masked, default linearity correction, 15$\sigma$ jump rejection threshold, and default ramp-fitting algorithm applied. In stage 2, the ramp images created in stage 1 ({\it rateints.fits}) are further minimally calibrated according to default \textsc{jwst} pipeline settings. We performed stage 2 reductions both with and without the photom step applied to test for any sensitivity in our solution to wavelength-dependent calibrations \citep[e.g., from the gain;][]{bell2023first}. The output of these first two stages are calibrated images ({\it calints.fits}) from which time-series stellar spectra can be extracted.

In stage 3, background subtraction and spectral extraction are performed on the calibrated integration images. Here, we selected a half-width aperture for the spectral extraction and background exclusion regions of four and ten pixels from the central pixel of the spectral trace, respectively, which served to minimize the scatter in our extracted spectra and avoided several problematic pixels. Background subtraction was performed row-by-row for each integration image with outlier identification performed along the time axis with a double-iteration 5$\sigma$ threshold rejection scheme. Optimal spectral extraction was then performed using a spatial profile defined by a median frame constructed from the time-series. A 5$\sigma$ threshold was applied in flagging outliers in the median frame and then a 10$\sigma$ threshold for outlier rejection was used for both constructing the spatial profile and during the optimal spectral extraction. In stage 4, spectroscopic and white-light light curves were then generated from the time series of the derived stellar spectra with a 5$\sigma$, five iteration clipping of outliers to a rolling temporal median spanning 10 data points.

\subsection{Light-curve Fitting and Spectral Generation}

Light-curve fitting was performed on the white-light curve (5--12 \textmu m), and at three spectroscopic resolutions with 0.125, 0.25, and 0.5 \textmu m wide bins for both the ExoTiC-MIRI and Eureka! reductions. These three different spectral resolutions were chosen to test the sensitivity of our derived spectra to the ``odd-even row effects" that are known to exist for the JWST MIRI detector \citep{ressler2015mid}. The light curves binned at 0.5 \textmu m are shown in Figure \ref{fig:light_curve_waterfall}. The independent ExoTiC-MIRI and Eureka! reductions both found that the light-curve model, $f(t)$, that best represented the astrophysical and systematic trends in the observations had the form
\begin{equation}
f(t) = T(t_{\rm{s}}, \mathbf{\theta}) \times S(t_{\rm{s}}, t_{\rm{c}}),
	\label{eq:exotic_miri_light_curve_model}
\end{equation}
where $T$ is the physical transit model, with the parameter vector $\mathbf{\theta}$, and the systematics model is described by
\begin{equation}
S(t_{\rm{s}}, t_{\rm{c}}) = r_0 \exp(r_1 t_{\rm{s}}) \times (c_0 + c_1 t_{\rm{c}}).
	\label{eq:exotic_miri_light_curve_systematics}
\end{equation}
Here, $r_0$, $r_1$, $c_0$, and $c_1$ are constants to be fit, and $t_{\rm{s}}$ and $t_{\rm{c}}$ are the observation times minus the light curve start time and centre-of-transit time, respectively. 

Both the ExoTiC-MIRI and Eureka! pipelines employed batman \citep{kreidberg2015batman} to generate the transit model, $T$. The ExoTiC-MIRI light-curve fits used a fixed quadratic limb-darkening law where the coefficients for each bin are computed using ExoTiC-LD \citep{david_grant_2022_7437681} and Set One of the MPS-ATLAS stellar models \citep{kostogryz2022stellar, kostogryz2023mps}. The Eureka! light-curve fits also used quadratic limb-darkening but reparameterized according to \citet{Kipping2013}, and additionally these fits decorrelated against the measured position and width of the spectral trace.  Both the ExoTiC-MIRI and Eureka! light-curve fits also include an error multiplier, $\beta$, to capture possible error underestimation/overestimation, in particular due to the uncertainties in the detector gain \citep{bell2023first}. The ExoTiC-MIRI and Eureka! light-curve fits cut off the first 100 (47~minutes) and 65 (30~minutes) integrations, respectively, to alleviate the model from fitting the worst of the systematic ramp. Additionally, the ExoTiC-MIRI light-curve fits manually masked the first data point from each segment, 12 in total, due to anomalous background levels, along with any remaining 4-sigma outliers from a running median. These same 12 integrations are flagged and removed in stage 4 of the Eureka! pipeline.

The fitting was performed on both the ExoTiC-MIRI and Eureka! generated light curves with the Markov chain Monte Carlo (MCMC) algorithm, emcee \citep{foreman2013emcee}. In both the ExoTiC-MIRI and Eureka! light-curve fitting, the white light curve was first fit to constrain the best-fit values for the center-of-transit time. Additionally, in the ExoTiC-MIRI light-curve fitting the semi-major axis and inclination were estimated from the white light curve, while Eureka! assumed the values from \citet{Alderson2022}. These system parameter values were then held fixed in the spectroscopic light-curve fitting. See Table \ref{tab:system_and_fitting_params} for a complete list of system parameters and prior distributions used in the light-curve fitting. The transmission spectra ($\left[\mathrm{R_p}/\mathrm{R_*}\right]^2$) derived from each reduction are displayed in Figure \ref{fig:transmission_spectra}.

\subsection{\textsc{ExoTiC-MIRI} and \textsc{Eureka!} Intercomparison}
\label{sec:reduction_intercomparison}

 Our analysis with both the ExoTiC-MIRI and Eureka! pipelines produces white light curves with precision (standard deviation of the normalised residuals, SDNR) measured at 465 and 453 ppm, respectively. The slight (12 ppm) precision increase in the Eureka! light curves is driven by two choices in the reductions. First, the ExoTiC-MIRI reduction discarded the first 12 groups from the start of each ramp, decreasing the ramp-fitting's statistical power with the aim of increasing accuracy, whereas the Eureka! reduction only discarded the first group. Second, the Eureka! reduction utilised optimal extraction, rather than a box aperture, and this was found to improve the precision.

The transmission spectra from the ExoTiC-MIRI and Eureka! pipelines exhibit a consistent morphological structure, albeit with one primary difference. At wavelengths $< 8$\,\textmu m the Eureka! spectrum is systematically deeper. Further investigation showed the different linearity corrections lead to this discrepancy. The derived custom linearity correction in ExoTiC-MIRI (see Section \ref{sec:obs-red-exotic}) is smaller than the default correction for the same increase in observed data numbers, and this correction is amplifier dependent. An analysis of the group-level ramps revealed the default linearity correction led to larger deviations from a linear ramp relative to the custom linearity correction. This is true across all detector columns, but critically, these deviations were largest (by 58\%) on the dispersion axis where most of the flux is collected. These differences are most likely due to ExoTiC-MIRI's custom correction better taking into account the debiasing-induced BFE \citep{argyriou2023brighter} specific to these data, which predominately impacts the bright pixels at the centre of the dispersion axis. This analysis indicates that the ExoTiC-MIRI reduction is more reliable, and we choose to focus our atmospheric modeling primarily on these data.

We note that at wavelengths $> 8$\,\textmu m the ExoTiC-MIRI and Eureka! transit depths are consistent to better than one standard deviation across all three resolutions, and the residuals between the two reductions pass the Shapiro-Wilk test for normality \citep{shapiro1965analysis}. Any small stochastic differences arise from the choices to fit or fix the limb-darkening coefficients, semi-major axis, inclination, and the type of extraction aperture. We found the transmission spectra were insensitive to decisions about the limb-darkening law within each reduction.

\section{Atmospheric Modeling and Retrievals} \label{sec:interp}
We employed grids of both self-consistent equilibrium-chemistry atmospheric forward models and ``free-chemistry" atmospheric retrievals to interpret our mid-infrared transmission spectrum of WASP-17b. We perform our fits on the 0.25 \textmu m ExoTiC-MIRI reduction using the four different models detailed below. Additional tests are run at each stage to confirm our results. We considered the combination of our JWST MIRI LRS spectrum with the previously presented visible-to-near-IR transmission spectrum of WASP-17b presented in \citet{Alderson2022}. In our spectral fitting we have accounted for the presence of a clear positive offset (${\sim}100-200$\,ppm) in the transit depths measured with JWST MIRI LRS and those measured with HST + {\it Spitzer}, which is common for multi-instrument and multi-epoch observations of transiting exoplanets \citep[e.g.,][]{Benneke2019NatAs, IhKempton2021AJ}. A summary of the modeling inputs and results is provided in Tables \ref{tab:forward_model_params} and \ref{tab:retrieval_model_params}.

\begin{figure*}
\includegraphics[width=1.0\textwidth]{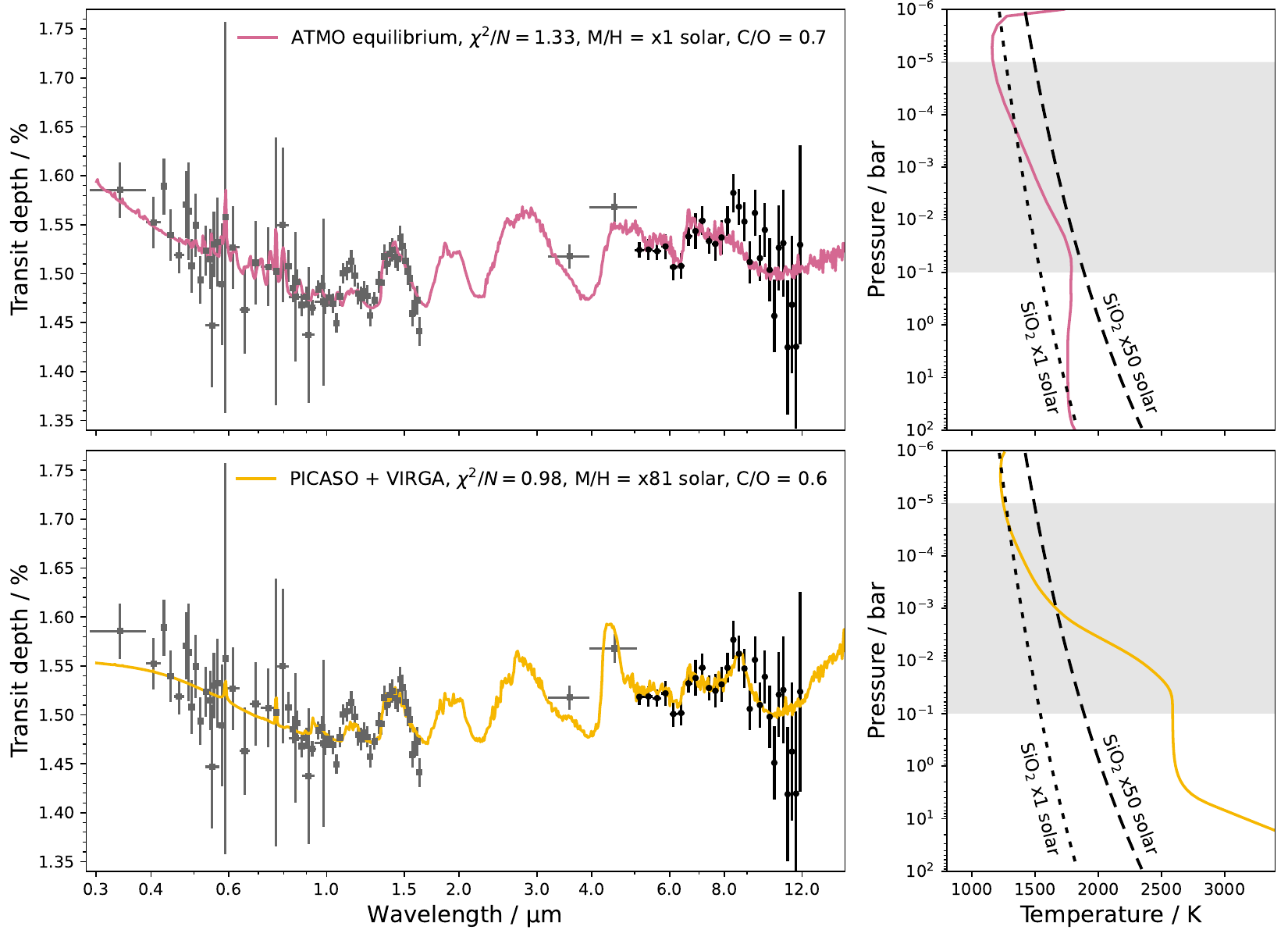}
\caption{Atmospheric modeling with ATMO (top panels, with a gray cloud and Rayleigh scattering haze prescription) and PICASO+Virga (bottom panels, with Mie-scattering cloud opacity of SiO$_2$(s) included). Here we show the best-fit transmission spectra applied to a combination of our JWST MIRI LRS data (0.25 \textmu m ExoTiC-MIRI reduction, black circles) and the \citet{Alderson2022} HST + {\it Spitzer} data (gray squares). The corresponding pressure-temperature profiles are shown in the right-hand panels, along with SiO$_2$'s condensation curves as defined in Equations \ref{eq:sio2_condensation_pressure} and \ref{eq:sio2_condensation_temperature}.}
\label{fig:forward_models_and_pt}
\end{figure*}

\begin{figure*}
\includegraphics[width=0.99\textwidth]{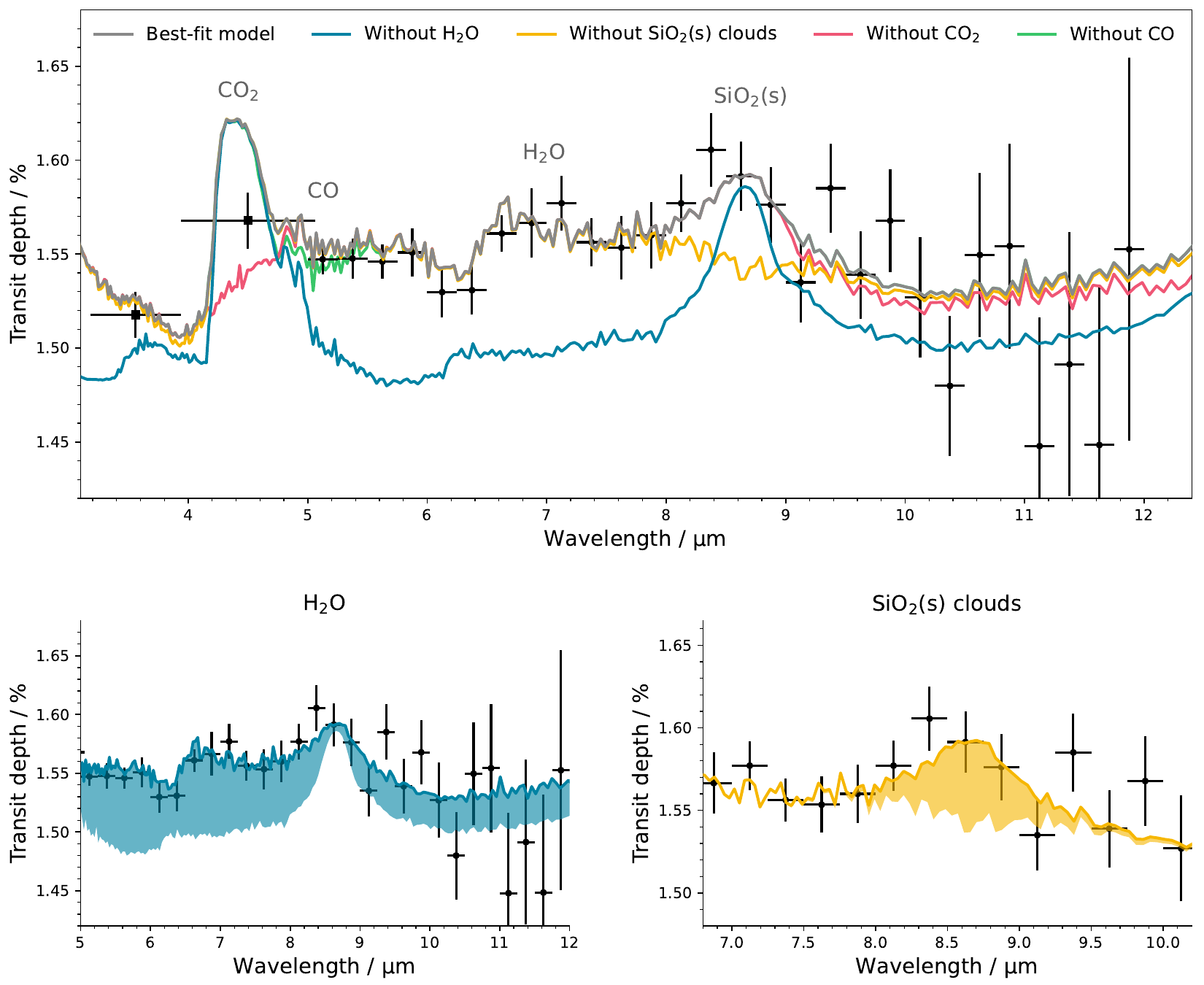}
\caption{Opacity contributions in our PICASO+Virga atmospheric model. Top panel: the best-fit model (gray) is shown against our JWST MIRI LRS data (0.25 \textmu m ExoTiC-MIRI reduction, black circles) and the \citet{Alderson2022} {\it Spitzer} data (black squares). We also show four additional versions of the same best-fit model, each having a different source of opacity removed, to reveal which features can be attributed to which atmospheric species. Bottom panels: zoom-in of the opacity contributions (shaded regions) from the two primary atmospheric species, H$_2$O and SiO$_2$(s) clouds, detected at mid-infrared wavelengths. A summary of these modeling results can be found in Table \ref{tab:forward_model_params}.}
\label{fig:opacity_contributions}
\end{figure*}

\subsection{\textsc{ATMO} Grid of Forward Models}
We used a planet-specific grid of self-consistent model atmospheres for WASP-17b generated using ATMO \citep{Tremblin2016, Amundsen2014, Drummond2016, Goyal2018}, with radiative-convective equilibrium pressure-temperature ($P$-$T$) profiles consistent with equilibrium chemistry \citep{Goyal2020}. This grid was generated for a range of heat redistribution factors (0.25, 0.5, 0.75, 1.0), metallicities (0.1x, 1x, 10x, 50x, 100x, 200x solar) and C/O ratios (0.35, 0.55, 0.70, 0.75, 1.0, 1.5). The internal temperature of the planet is assumed to be 100\,K, although this value is not well known (see e.g., \citet{thorngren2019intrinsic} and \citet{sarkis2021evidence}). We test internal temperatures of 100\,K, 200\,K, and 300\,K and find the differences in the modeled transmission spectrum to be negligible. Using this grid of WASP-17b model atmospheres, a grid of simulated transmission spectra was generated at $R \sim 1000$ for a range of Rayleigh scattering haze factors (1x, 10x nominal Rayleigh scattering) and gray cloud factors (0.0x, 0.5x, 1.0x, 5.0x H$_2$ Rayleigh scattering cross-section at 350 nm). 

We find that the best-fit ATMO model to the combined HST + {\it Spitzer} and JWST MIRI LRS datasets has a re-distribution factor of 0.5, solar metallicity, super-solar C/O ratio (0.7), haze factor of 10, and gray cloud factor of 0.5. This model is shown in the top panels of Figure \ref{fig:forward_models_and_pt}. We note that models with 0.1x to 50x solar metallicity and 0.35 to 0.7 C/O ratios lie within the 3$\sigma$ range of the best-fit model. We found consistent results across all 3 binning regimes for both the ExoTiC-MIRI and the Eureka! reductions presented in Figure \ref{fig:transmission_spectra}, but note that the best-fit offsets between the JWST MIRI LRS and HST + {\it Spitzer} data differed between ExoTiC-MIRI and Eureka! pipeline reductions. As previously discussed, the different linearity corrections applied in the ExoTiC-MIRI and Eureka! reductions resulted in offsets in the transit depth shortward of 8\,\textmu m. However, we find that allowing for an offset between the JWST MIRI LRS data and the HST + {\it Spitzer} data results in identical best-fitting atmospheric models, which indicates that our atmospheric interpretation is minimally sensitive to differences between the ExoTiC-MIRI and Eureka! reductions or the choice of spectral bin size.

\subsection{\textsc{PICASO+Virga} Grid of Forward Models}

We computed radiative-convective thermochemical equilibrium (RCTE) atmospheric models for WASP-17b using the well-vetted open-source model PICASO v3.1\footnote{\url{https://github.com/natashabatalha/picaso}} \citep{batalha2019exoplanet, mukherjee2023picaso}, which has heritage from Fortran codes developed to study Solar System giant planets \citep[e.g.,][]{marley1999thermal} and brown dwarfs \citep[e.g.,][]{marley1999thermal}.
We computed a grid of cloud-free models as a function of interior temperature of the planet (200 K \& 300 K, \citealt{thorngren2019intrinsic, sarkis2021evidence}), atmospheric metallicity (9 values between 1--100$\times$Solar), C/O ratio (5 values between 0.25--2$\times$Solar), and the heat redistribution factor (0.5, 0.6, 0.7, 0.8). PICASO’s RCTE module utilizes correlated-k opacities that are detailed in \citet{Marley2021} and released by \citet{lupu_roxana_2021_5590989}. We included the opacity sources for 29 species, but the most important for the JWST MIRI LRS wavelength region is the line list of H$_2$O \citep{Polyansky2018H2O}. The chemical equilibrium abundances are computed on a pressure-temperature-M/H-C/O grid of thermochemical equilibrium models presented in \citet{Marley2021} following the work of \citet{gordon1994computer}, \citet{fegleylodders1994}, \citet{lodders99}, \citet{lodders02}, \citet{LoddersFegley2002}, \citet{visscher06}, and \citet{channon10}, using elemental abundances from \citet{Lodders2010}. From our WASP-17b climate models we computed transmission spectra using opacities resampled to $R = 60,000$ \citep{natasha_batalha_2020_6928501} from original $R \sim 10^6$ line-by-line calculations detailed in \citet{freedman2008opacities} and \citet{nezhad21}. We fit the cloud-free grid to both the ExoTiC-MIRI and Eureka! JWST MIRI LRS reductions using the ``MLFriends'' nested sampling Algorithm \citep{MLFriends2016, MLFriends2019} implemented in the open-source Ultranest code \citep{Ultranest} and find agreement between the two, regardless of binning scheme. The best-fit model from the cloud-free grid has an internal temperature of 200~K, redistribution factor of 0.8, metallicity of 100$\times$Solar, and super-solar C/O ratio (0.7). 

In order to account for the presence of clouds in WASP-17b's atmosphere, we used the cloud-free grid's pressure-temperature profiles to compute condensation cloud profiles using the Virga cloud model\footnote{\url{https://github.com/natashabatalha/virga}} \citep{Batalha2020, Rooney2022}. The cloud methodology of Virga is detailed in \citet{Ackerman-Marley2001}. Virga models the balance between the turbulent diffusion (K$_{zz}$) and sedimentation (f$_\mathrm{sed}$) in horizontally uniform cloud decks, and therefore requires these two additional model parameters. We include two condensable species, SiO$_2$(s) and Al$_2$O$_3$(s), given the range of temperatures expected in WASP-17b's atmosphere and presence of a clear spectroscopic feature near 8.6\,\textmu m. For SiO$_2$(s), we used the $\alpha$-crystal optical constants computed at 928 K by \citet{zeidler2013optical}. For Al$_2$O$_3$(s), we used the amorphous optical constants computed at 873 K by \citep{koike1995extinction, begemann1997aluminum}. We note that we did explore a number of other cloud species, including MgSiO$_3$(s) and Mg$_2$SiO$_4$(s), for which we show the cross-sections in the left-hand panel of Figure \ref{fig:cross_sections}. But, these cloud species could not match the observed mid-infrared spectroscopic features. We also tested the amorphous form of SiO$_2$(s), but this results in a worse match to the 8.6\,\textmu m feature, and the reduced $\chi^2$ of the fit increases from 0.98 to 1.05 (90 degrees of freedom).

The saturation vapor pressure curve of Al$_2$O$_3$ was calculated in \citet{wakeford2016high}. For SiO$_2$, we calculated the vapor pressure and condensation curve for the net thermochemical reaction, SiO(g) + H2O(g) = SiO$_2$(s) + H2(g), which is adopted because SiO and H$_2$O are the dominant Si- and O-bearing gases, respectively, at the temperatures pertinent to WASP-17b. This leads to condensation curves given by 
\begin{equation}
    \log P(\mathrm{SiO}) \approx 13.168 \,-\, 28265 / T \,-\, \mathrm{[Fe/H]},
    \label{eq:sio2_condensation_pressure}
\end{equation}
or equivalently,
\begin{equation}
10^4/T_{\rm{cond}}(\mathrm{SiO_2}) \approx 6.14 \,-\, 0.35 \log P_T \,-\, 0.70 \mathrm{[Fe/H]},
    \label{eq:sio2_condensation_temperature}
\end{equation}
where $P_T$ is in bars. These condensation curves are shown in the right-hand panels of Figure \ref{fig:forward_models_and_pt}.

We fit our cloudy WASP-17b spectra to the ExoTiC-MIRI reduction combined with HST + {\it Spitzer} data from \citet{Alderson2022} using Ultranest \citep{Ultranest}. For the parameters confined to the grid we obtain 1$\sigma$ constraints of 30-100$\times$Solar metallicity, 0.4-0.7 C/O, a heat redistribution of 0.6-0.7, and internal temperature of 200-300~K. For the two cloud parameters the 1$\sigma$ constraints are $\log K_{zz}=9^{+0.75}_{-0.29}$\,cm$^2$/s and $f_{\rm{sed}} = 0.1^{+0.22}_{-0.76}$. We compared the cloudy model with the cloud-free model using their likelihood ratio, and we find that the SiO$_2$(s) cloud model is preferred by 4.2$\sigma$. The maximum-likelihood cloudy model is shown in the bottom panels of Figure \ref{fig:forward_models_and_pt}. Additionally, in Figure \ref{fig:opacity_contributions} the opacity contributions to this model are shown, highlighting that H$_2$O vapor and SiO$_2$(s) clouds dominate WASP-17b's transmission spectrum at JWST MIRI LRS wavelengths.

\subsection{\textsc{POSEIDON} Atmospheric Retrievals}

We conducted free-chemistry retrievals with the inclusion of compositionally specific aerosols on HST, {\it Spitzer}, and JWST MIRI LRS observations of the transmission spectrum of WASP-17b using the open-source atmospheric retrieval code POSEIDON\footnote{\url{https://github.com/MartianColonist/POSEIDON}} \citep{MacDonaldMadhusudhan2017, MacDonald2023}. Here we build upon the existing cloud models in POSEIDON to include retrievals with Mie-scattering aerosols. We calculated a database of effective extinction cross sections of aerosols species by adapting \citet{Zhang2019}'s implementation of the Mie-scattering algorithm presented in \citet{Kitzmann2018}. The effective aerosol extinction cross section is the combined absorption and scattering cross section integrated over a log-normal radius distribution centered around a mean particle size $r_m$ (\textmu m). For computational efficiency, we precomputed effective extinction cross sections from the refractive-index databases of \cite{Wakeford2015} and \cite{Kitzmann2018}, where the relevant species are based on work by \citet{henning1997low}, \citet{palik1998handbook}, \citet{andersen2006infrared}, and \citet{zeidler2013optical}. We considered mean particle sizes ranging from 0.001--10\,\textmu m and wavelengths spanning 0.2--30\,\textmu m at $R=1000$.

Our POSEIDON retrievals adopt a model configuration similar to \citet{Alderson2022}, but with the addition of a parameterized Mie-scattering cloud model. We assume a one-dimensional H$_2$-He dominated atmosphere (with He/H$_2$ = 0.17) with the gas-phase containing free abundances of H$_2$O, CH$_4$, CO$_2$, CO, Na, and K. The specific line lists used are detailed in the appendix of \citet{macdonald2022trident}, with particular reference to \citet{Polyansky2018H2O}, \citet{yurchenko2017hybrid}, \citet{chubb2021exomolop}, \citet{tashkun2011cdsd}, \citet{li2015rovibrational}, and \citet{barklem2016partition}. We assume an isothermal pressure-temperature profile, which follows from the modeling of \citet{Alderson2022}. The reference pressure is set at 10\,bar. We ran retrievals both with and without the JWST MIRI LRS data (0.25 \textmu m ExoTiC-MIRI reduction), but always including the HST and Spitzer data. We consider three models: (1) a cloud-free model; (2) the gray cloud+scattering haze prescription found in \citet{Alderson2022}; and (3) Mie-scattering SiO$_2$(s) clouds (both crystalline and amorphous states considered separately). Aerosol clouds are parameterized by the mean particle size, $r_{\rm{m}}$, the cloud-top pressure, $P_{\rm{cloud}}$, the width of the cloud in log pressure space, $\Delta\log$\,$P$, and the constant log mixing ratio of the aerosol in the cloud. The priors for our POSEIDON retrievals are summarized in Table \ref{tab:retrieval_model_params}.

We computed the marginalised likelihood for each of the models and compared them via the Bayes factor to assess the support for each model. We find that the SiO$_2$(s) clouds model is preferred over no clouds by 3.5$\sigma$ and also preferred over the generic aerosol prescription by 2.6$\sigma$. The best-fit model finds the SiO$_2$(s) clouds are comprised of small particles ($\sim$ 0.01\,\textmu m) with a low mixing ratio ($\log{X} \sim -12$) located high in the atmosphere ($P_{\rm{cloud}} \lesssim$ 1\,mbar). Our data are highly informative of the particle size, as changes to this parameter dramatically alter both the optical scattering slope and the amplitude of the 8.6\,\textmu m vibrational-mode feature (see Figure \ref{fig:particle_sizes}). Both crystalline and amorphous SiO$_2$(s) aerosols are able to produce good fits to the data, with the crystalline form producing the best match to the peak position of the feature near 8.6\,\textmu m. In addition to SiO$_2$(s), we also considered Fe$_2$O$_3$(s) aerosols based on the alignment of refractive indices with the observed feature near 8.6\,\textmu m (see Figure \ref{fig:cross_sections}). However, the best-fit model requires particle sizes of $\sim$ 0.001\,\textmu m, which is physically implausible \citep{sridevi2023Fe2O3structural}. Thus we rule out Fe$_2$O$_3$(s) aerosols.

\begin{figure*}
\includegraphics[trim={0 1.2cm 0 0},clip,width=0.91\textwidth]{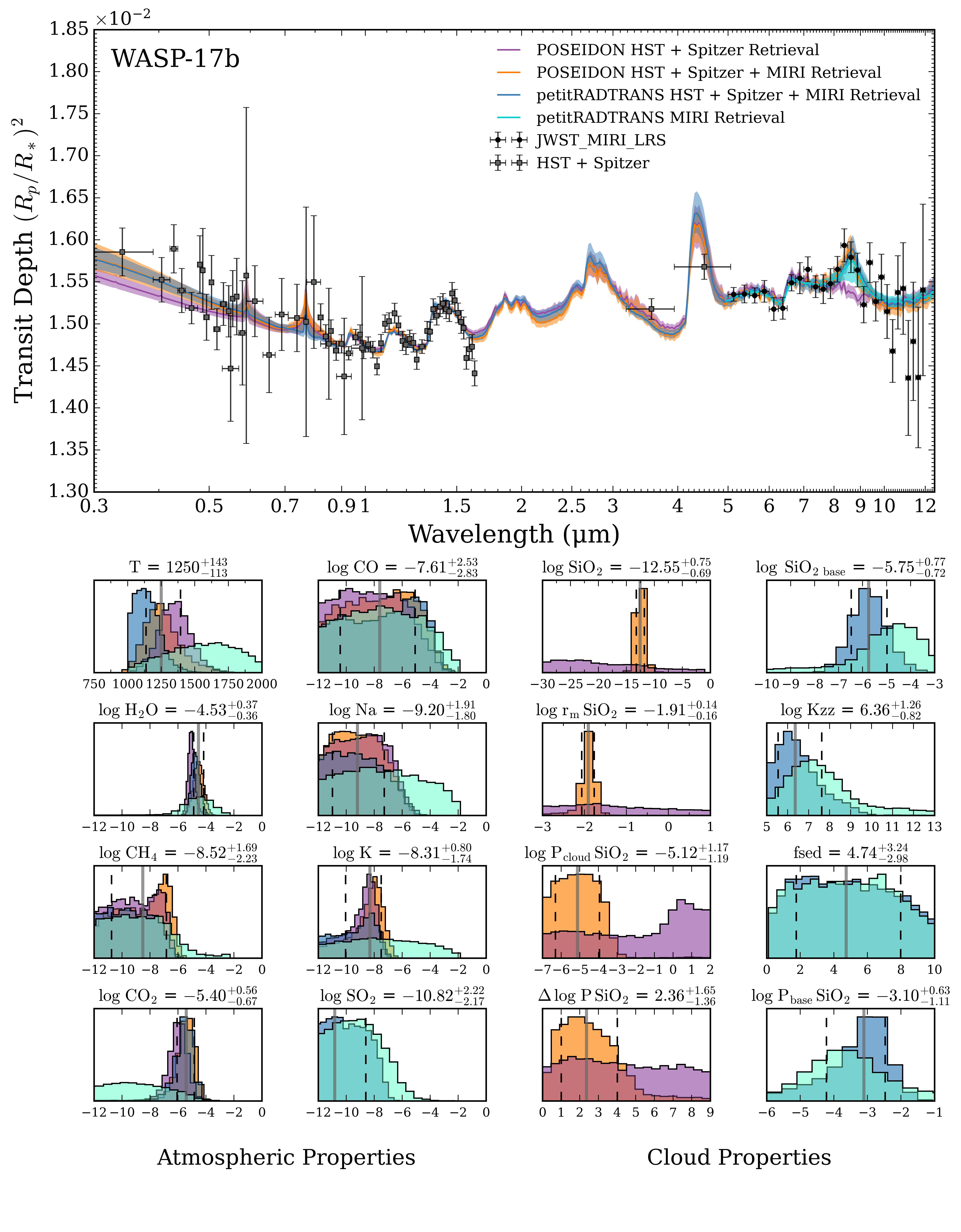}
\caption{Atmospheric retrievals with POSEIDON and petitRADTRANS using combinations of the JWST MIRI LRS (0.25 \textmu m ExoTiC-MIRI reduction) and the \citet{Alderson2022} HST + {\it Spitzer} data (see legend in the top panel). Top panel: median retrieved spectra (solid lines) and $1\sigma$ confidence regions (shaded regions). Bottom panels: retrieved posteriors for isothermal atmospheric temperature and gas-phase volume mixing ratios (columns one and two) and cloud properties from POSEIDON (column three) and petitRADTRANS (column four). POSEIDON's agnostic cloud model is described by a constant aerosol mixing ratio, $\log$\,SiO$_2$, mean particle size, $\log$\,$r_{\rm{m}}$\,SiO$_2$, cloud-top pressure, $\log$\,$P_{\rm{cloud}}$\,SiO$_2$, and width of the cloud, $\Delta\log$\,$P$\,SiO$_2$. petitRADTRAN's physically motivated cloud model is described by the aerosol mixing ratio at the cloud base, $\log$\,SiO$_{2\,\rm{base}}$, vertical mixing, $\log$\,$K_{zz}$, sedimentation efficiency, $f_{\rm{sed}}$, and the cloud-base pressure, $\log$\,$P_{\rm{base}}$\,SiO$_2$. A summary of these modeling results can be found in Table \ref{tab:retrieval_model_params}.}
\label{fig:Retrievals}
\end{figure*}

For completeness, we also retrieve on our 0.25 \textmu m Eureka! reduction. These retrievals also yield a preference for clouds, with both the SiO$_2$(s) and generic clouds preferred at 2.6$\sigma$ and 3.0$\sigma$, respectively, over the cloud-free model. While the SiO$_2$(s) model produces a better fit to the Eureka! data than the generic clouds (reduced $\chi^2$ = 1.14 versus 1.17), the Bayes factor does not support the SiO$_2$(s) model given the additional parameters in its definition. However, as detailed in section \ref{sec:reduction_intercomparison}, the Eureka! reduction may be less reliable than the ExoTiC-MIRI reduction at shorter wavelengths and this impacts the amplitude and detectability of the 8.6\,\textmu m feature.

Retrievals that do not include JWST MIRI LRS data indicate no preference over either of our cloud models, a result consistent with predictions from \citet{Mai2019}. In all cases, our retrieved abundances for the gas-phase chemical species considered are consistent with the values presented in \citet{Alderson2022}. In Figure \ref{fig:Retrievals} we show retrieved spectra with 1$\sigma$ contours and posterior distributions for gas-phase species (H$_2$O, CH$_4$, CO$_2$, Na, K, CO), SiO$_2$(s) cloud properties, and isothermal atmospheric temperature.

\subsection{\textsc{petitRADTRANS} Atmospheric Retrievals}

To complement our POSEIDON retrievals, we performed free retrievals using petitRADTRANS\footnote{\url{https://gitlab.com/mauricemolli/petitRADTRANS}} (pRT) \citep{molliere2019petitradtrans}, which has been used previously to explore specific aerosols and chemistry in brown dwarfs and directly imaged exoplanet atmospheres \citep[e.g.,][]{Molliere2020}. Similar to POSEIDON, we opt to use an isothermal prescription for the pressure-temperature profile and consider the case of a H$_2$-He-dominated atmosphere with H$_2$O, CO$_2$, SO$_2$, CH$_4$, CO, Na, and K as possible gas-phase species. The specific line lists used are detailed in \citet{molliere2019petitradtrans}, with particular reference to \citet{rothman2010hitemp}, \citet{yurchenko2017hybrid}, \citet{chubb2021exomolop}, \citet{underwood2016exomol}, \citet{tobias2018critical}, and \citet{piskunov1995vald}. The reference pressure is set at 100\,bar. We also allow for the presence of crystalline SiO$_2$(s) and use the same refractive indices as POSEIDON. However, we leverage the cloud model implementation built into pRT which is based on \citet{Ackerman-Marley2001}. This cloud model includes parameters for vertical mixing, $K_{zz}$, sedimentation efficiency, $f_{\rm{sed}}$, and cloud-base pressure, $P_{\rm{base}}$, that give rise to a population of cloud particles with a log-normal size distribution and a specific mixing ratio across a range of pressures above the cloud base. The priors for our pRT retrievals are also summarized in Table \ref{tab:retrieval_model_params}.

We performed pRT retrievals on the JWST MIRI LRS data alone (0.25 \textmu m ExoTiC-MIRI reduction) and in combination with the HST + {\it Spitzer} data presented in \citet{Alderson2022}. The results are displayed in Figure \ref{fig:Retrievals} alongside those from POSEIDON. We find abundances for the gas-phase species considered are consistent with estimates from POSEIDON, noting that only H$_2$O is constrained from the JWST MIRI LRS data alone and there is no strong evidence for SO$_2$, a photochemically produced species that has absorption signatures at MIRI wavelengths \citep{Tsai2023}. In our pRT retrievals, we find that the inclusion of SiO$_2$(s) clouds is preferred over a generic-clouds model, with $\chi^2$ values of 100.0 and 129.9 (80 and 82 degrees of freedom), respectively. The SiO$_2$(s) clouds have a cloud base in the 1-10\,mbar region of WASP-17b's limb, and an average particle size of ${\sim}$0.01\,\textmu m, which is consistent with the findings from POSEIDON. These clouds are produced within the pRT cloud model with $K_{zz}\sim 10^6 \,\rm{cm}^{2}\rm{s}^{-1}$ and an $f_{sed}$ on the order of 5, which is consistent with the best-fit models from PICASO+Virga considering the trade-offs between these two parameters in the context of the \citet{Ackerman-Marley2001} cloud model.

\subsection{Atmospheric Modeling Intercomparison}
Our suite of forward models and retrievals for WASP-17b have produced evidence for an atmosphere shaped by complex chemistry, where aerosols are required to fit the transmission spectrum of WASP-17b. Across all of our modeling efforts, we have trialled aerosols composed of SiO$_2$(s), Al$_2$O$_3$(s), Fe$_2$O$_3$(s), MgSiO$_3$(s), and Mg$_2$SiO$_4$(s). We find that SiO$_2$(s) is favoured as the dominant cloud species, and is statistically preferred over no clouds by 3.5$\sigma$ in our POSEIDON free retrievals and at 4.2$\sigma$ in our PICASO equilibrium-chemistry forward models. We additionally investigated a combination of SiO$_2$(s) and generic aerosol parameterisations with POSEIDON, and find that SiO$_2$(s) is still required to fit the spectrum at the 2.6$\sigma$ level. This modeling suggests that small-particle SiO$_2$(s) clouds are present at the limb of WASP-17b's atmosphere.

Our free-chemistry retrievals all find an overall sub-solar abundance of H$_2$O, while our equilibrium-chemistry forward models have preferences for super-solar metallicities and C/O ratios. These differences are most likely caused by the constraints imposed by RCTE in the forward models. In particular, the PICASO+Virga cloudy model has its metallicity driven to higher values because in Virga the gas mean mass mixing ratio is fixed to a constant value. Therefore, higher-metallicity solutions may be needed to compensate for the low cloud opacity. The difference between the super-solar metallicity derived from both equilibrium-chemistry forward models and the sub-solar metallicities found by the free-chemistry retrievals suggest that disequilibrium processes could be at play in WASP-17b's atmosphere.

All of our atmospheric models demonstrate how the mid-IR observations are key to identifying the cloud species, while observations at optical wavelengths may act to further constrain the cloud's average grain size, density, and vertical location. Overall, our ``four-pronged" theoretical exploration of WASP-17b's atmospheric thermal structure and composition all point to a planetary transmission spectrum strongly shaped by H$_2$O vapor and SiO$_2$(s) clouds, with suggestions of absorption from CO and CO$_2$ that will be more fully explored with our JWST NIRISS and NIRSpec observations of this planet (GTO-1353).

\section{Discussion and Conclusions} \label{sec:discuss_conclusions}
We have presented a first look at the atmosphere of the hot Jupiter WASP-17b in the mid-infrared with JWST MIRI LRS from 5--12\,\textmu m. We performed multiple reductions of the data and demonstrated the importance of understanding the nuances of the MIRI detector. We found that different approaches to the linearity correction can lead to systematic differences at wavelengths $< 8$\,\textmu m. Our analysis of the group-level ramps showed that a self-calibrated linearity correction may be the best approach for accounting for the debiasing-induced BFE \citep{argyriou2023brighter}, and obtaining reliable spectra.

The transmission spectrum suggests evidence for the specific absorption signature of SiO$_2$(s) clouds, with a peak in the transmission spectrum centered at 8.6\,\textmu m. When including the HST + {\it Spitzer} data, this broad wavelength coverage enabled us to constrain the size of the SiO$_2$(s) particles (${\sim}0.01$\,\textmu m) through fitting the optical scattering slope and the amplitude of the 8.6\,\textmu m vibrational-mode feature, simultaneously. The SiO$_2$(s) clouds model is preferred at 3.5--4.2$\sigma$ versus a cloud-free model and at 2.6$\sigma$ versus a generic aerosol prescription. The shape of the feature is also likely modified from that of pure SiO$_2$(s) when combined with opacity from H$_2$O, as shown in Figure \ref{fig:opacity_contributions}, or by aerosol aggregates composed of multiple species (dirty grains). Based on the available optical constants, we found that crystalline SiO$_2$(s) provides a better match to the peak position of the feature near 8.6\,\textmu m compared with amorphous SiO$_2$(s), but we note that for both species these have not been measured under WASP-17b conditions and further lab work is needed in the exoplanet regime. 

Atmospheric modeling of the transmission spectrum with equilibrium-chemistry forward models and free-chemistry retrievals found a super-solar metallicity (up to 100$\times$ solar) with a depletion in H$_2$O and significant opacity from aerosols. Previous work by \citet{Alderson2022} showed a bimodal (sub-solar and super-solar) solution for H$_2$O based on the measured HST + {\it Spitzer} spectrum. We find that the addition of the JWST MIRI LRS wavelength range serves to solidify the finding of a sub-solar abundance of H$_2$O in WASP-17b's atmosphere. The depletion of H$_2$O and super-solar C/O in the gas phase are likely a direct result of the formation of high-temperature aerosols which is dominated by oxygen-rich species. This fits well with theoretical studies which suggest that up to 30\% of oxygen can be depleted from the gas-phase chemistry when condensate cloud formation is dominant in the atmosphere of hot Jupiters similar to WASP-17b \citep[e.g.,][]{Lee2016,Lines2018}. 

\citet{Helling2006} first suggested that SiO$_2$(s) should be the most abundant solid component in L-type brown dwarf ``dust clouds" under non-equilibrium conditions and would give rise to strong absorption features. Infrared observations and atmospheric retrievals of brown dwarfs have shown hints of the presence of SiO$_2$(s) grains in their atmospheres \citep{Burningham2021}, but nothing to match the signature that we see in WASP-17b's transmission spectrum. \citet{Burningham2021} also found that SiO$_2$ may condense (instead of forsterite, but alongside enstatite) in atmospheres with subsolar Mg/Si abundance ratios. In studies on brown dwarfs, \citet{Helling2006} and \citet{HellingWoitke2006} suggested that seed particles of TiO$_2$(s) would allow for heterogeneous formation of ``dirty grains'' in the upper atmosphere. For hot Jupiters like WASP-17b, it is likely that horizontal transport of cloud particles across significant gradients in temperature could govern a similar heterogeneous formation process. The expected horizontal temperature range in WASP-17b's atmosphere \citep[e.g.,][]{kataria2016atmospheric, zamyatina2023observability} is well aligned with the L-dwarf scenarios explored by \citet{HellingWoitke2006} that resulted in the production of a population of small (${\sim}10^{-2}$~\textmu m) grains with a significant fraction of SiO$_2$(s) at pressures relevant to transmission spectroscopy (1-10~mbar). This strong indication of spectroscopic features, indicative of heterogeneous cloud formation processes in WASP-17b, opens up many new avenues for observational, theoretical, and laboratory exploration of aerosol formation and transport in exoplanetary atmospheres that will be vital for future measurements. 

\vspace{0.7cm}
We thank the anonymous referee for a helpful and constructive report. This paper reports work carried out in the context of the JWST Telescope Scientist Team (PI: Mountain). Funding is provided to the team by NASA through grant 80NSSC20K0586. Observations with the NASA/ESA/CSA JWST are associated with program GTO-1353 (PI: Lewis), obtained at the Space Telescope Science Institute, which is operated by AURA, Inc., under NASA contract NAS 5-03127. The JWST data presented in this paper were obtained from the Mikulski Archive for Space Telescopes (MAST) at the Space Telescope Science Institute. The specific observations analyzed can be accessed via \dataset[10.17909/e61r-hk80]{https://doi.org/10.17909/e61r-hk80}, and data products and models are available at \url{https://doi.org/10.5281/zenodo.8360121}.

D.G. acknowledges funding from the UKRI STFC Consolidated Grant ST/V000454/1. H.R.W. was funded by UK Research and Innovation (UKRI) under the UK government’s Horizon Europe funding guarantee [grant number EP/Y006313/1]. A.G. acknowledges support from the Robert R. Shrock Graduate Fellowship. JG acknowledges funding from SERB Research Grant SRG/2022/000727. R.J.M. is supported by NASA through the NASA Hubble Fellowship grant HST-HF2-51513.001, awarded by the Space Telescope Science Institute, which is operated by the Association of Universities for Research in Astronomy, Inc., for NASA, under contract NAS 5-26555. We acknowledge the MIT SuperCloud and Lincoln Laboratory Supercomputing Center for providing high performance computing resources that have contributed to the research results reported within this paper.  Resources supporting this work were provided by the NASA High-End Computing (HEC) Program through the NASA Advanced Supercomputing (NAS) Division at Ames Research Center. N.E.B. acknowledges support from NASA’S Interdisciplinary Consortia for Astrobiology Research (NNH19ZDA001N-ICAR) under award number 19-ICAR19\_2-0041. We thank Evert Nasedkin for assistance using the pRT retrieval code.

\vspace{5mm}
\facilities{JWST(MIRI LRS), HST(WFC3, STIS), {\it Spitzer}(IRAC)}

\software{
ExoTiC-MIRI \citep{grant_david_2023_8211207},
Eureka! \citep{Bell2022}, 
ExoTiC-LD \citep{david_grant_2022_7437681},
batman \citep{kreidberg2015batman},
emcee \citep{foreman2013emcee},
ATMO \citep{Tremblin2016, Amundsen2014, Drummond2016, Goyal2018}, 
PICASO \citep{batalha2019exoplanet, mukherjee2023picaso}, 
Virga \citep{Batalha2020, Rooney2022}, 
POSEIDON \citep{MacDonaldMadhusudhan2017, MacDonald2023},
petitRADTRANS \citep{molliere2019petitradtrans},
numpy \citep[][]{harris2020array}, SciPy \citep[][]{2020SciPy-NMeth}, matplotlib \citep[][]{Hunter:2007}, xarray \citep{hoyer2017xarray, hoyer_stephan_2022_6323468}, astropy \citep{2013A&A...558A..33A,2018AJ....156..123A, price2022astropy}.
}

\newpage
\appendix

\begin{deluxetable*}{l c c c c}[h!]
    \renewcommand{\arraystretch}{1.1}
    \tabletypesize{\footnotesize}
    \tablecolumns{5} 
    \tablecaption{Light curve fitting parameter information. Values are shown for the white light curve fits.}
    \tablehead{ & \multicolumn{2}{c}{ExoTiC-MIRI} & \multicolumn{2}{c}{Eureka!}}
    \startdata
    Parameter & Prior & Value & Prior & Value \\
    \hline
    $P$ [days] & fixed \citep{Alderson2022} & $3.73548546$ & fixed \citep{Alderson2022} & $3.73548546$ \\
    $t_0$ [$\mathrm{BJD}_\mathrm{TDB} - 2460016.7$] & $\mathcal{U}$(-0.1, 0.1) & $0.026452 \pm 0.000061$ & $\mathcal{N}$(0.02648, 0.05) &  $0.026477 \pm 0.000060$ \\
    $a/R_{*}$ & $\mathcal{U}$(6, 8) & $7.110 \pm 0.040$ & fixed \citep{sedaghati2016potassium} & $7.025$ \\
    $i$ [degrees] & $\mathcal{U}$(80, 90) & $87.217 \pm 0.135$ & fixed \citep{Alderson2022} & $86.9$ \\
    $\mathrm{R_p}/\mathrm{R_*}$ & $\mathcal{U}$(0.10, 0.15) & $0.12472 \pm 0.00016$ & $\mathcal{N}$(0.123, 0.05)  & $0.12506 \pm 0.00019$ \\
    $T_{\rm{eff}}$ [K] & fixed \citep{southworth2012homogeneous} & 6550 & $\cdots$ & $\cdots$ \\
    Fe/H & fixed \citep{southworth2012homogeneous} & -0.25 & $\cdots$ & $\cdots$ \\
    $\log g$ [cm/s$^2$] & fixed \citep{southworth2012homogeneous} & 4.149 & $\cdots$ & $\cdots$ \\
    $u_1$ & fixed & $0.05$ & $\cdots$ & $\cdots$ \\
    $u_2$ & fixed & $0.06$ & $\cdots$ & $\cdots$ \\
    $q_1$ & $\cdots$ & $\cdots$ & $\mathcal{U}$(0, 0.05) &  $0.0067 \pm 0.0016$\\
    $q_2$ & $\cdots$ & $\cdots$ & $\mathcal{U}$(0.15, 0.35) &  $0.258 \pm 0.069$\\
    \hline
    $c_0$ & $\mathcal{N}$(1, 0.01) & $1.000091 \pm 0.000035$ & $\mathcal{N}$(1.001, 0.01) & $1.006164 \pm 0.000055$ \\
    $c_1$ & $\mathcal{N}$(0, 0.01) & $-0.00033 \pm 0.00026$ & $\mathcal{N}$(0, 0.01) & $-0.00102 \pm 0.00036$ \\
    $r_0$ & $\mathcal{N}$(0, 0.1) & $0.00135 \pm 0.00011$ & $\mathcal{N}$(0, 0.01)  & $0.00151 \pm 0.00022$ \\
    $r_1$ & $\mathcal{U}$(0, 200) & $40.3 \pm 8.0$ & $\mathcal{U}$(5, 150)  & $27.5 \pm 5.1$ \\
    $\beta$ & $\mathcal{U}$(0.1, 5) & $1.16 \pm 0.02$ & $\mathcal{N}$(1.5, 0.5)  &  $1.724 \pm 0.035$ \\
    $x_0$ & $\cdots$ & $\cdots$ & $\mathcal{N}$(0, 0.1) &  $0.0439 \pm 0.0069$\\
    $x_1$ & $\cdots$ & $\cdots$ & $\mathcal{N}$(0, 0.5) &  $0.026 \pm 0.020$\\
    \enddata
    \label{tab:system_and_fitting_params}
    \tablecomments{Parameter definitions: orbital period, $P$; time of transit centre, $t_0$; semi-major axis in units of stellar radius, $a/R_{*}$; inclination, $i$; planet radius in units of stellar radius, $\mathrm{R_p}/\mathrm{R_*}$; stellar effective temperature, $T_{\rm{eff}}$; stellar metallicity, Fe/H; stellar gravity, $\log g$; quadratic limb-darkening coefficients, $u_1$ and $u_2$; \citet{Kipping2013} reparameterisation of the quadratic limb-darkening coefficients, $q_1$ and $q_2$, systematic model parameters as defined in Equation \ref{eq:exotic_miri_light_curve_systematics}, $c_0$, $c_1$, $r_0$, $r_1$; error multiplier, $\beta$; decorrelation coefficients for trace position and width, $x_0$ and $x_1$. Units for logarithmic parameters refer to the argument.}
\end{deluxetable*}

\begin{deluxetable*}{l c c c c}
    \renewcommand{\arraystretch}{1.1}
    \tabletypesize{\footnotesize}
    \tablecolumns{5} 
    \tablecaption{Summary of atmospheric forward modeling.}
    \tablehead{ & \multicolumn{2}{c}{\textsc{Atmo}} & \multicolumn{2}{c}{\textsc{PICASO+Virga}}}
    \startdata
    Parameter\phantom{gap} & \phantom{gap}Range\phantom{gap} & \phantom{gap}Best-fit value\phantom{gap} & \phantom{gap}Range\phantom{gap} & \phantom{gap}Fitted 1$\sigma$ range\phantom{gap} \\
    \hline
    Metallicity & 0.1--200 & 1 & 1--100 & 30--100 \\
    C/O & 0.35--1.5 & 0.7 & 0.1--0.9 & 0.4--0.7 \\
    Heat redistribution & 0.25--1.0 & 0.5 & 0.5--0.8 & 0.6--0.7 \\
    Internal temperature [K] & fixed & 100 & 200--300 &  200--300 \\
    Gray cloud factor & 0--5 & 0.5 & $\cdots$ & $\cdots$\\
    Rayleigh scattering haze factor & 1--10 & 10 & $\cdots$ & $\cdots$ \\
    $\log K_{zz}$ [cm$^2$/s] & $\cdots$ & $\cdots$ &  & 8.7--9.8 \\
    $f_{\rm{sed}}$ & $\cdots$ & $\cdots$ &  & -0.7--0.3 \\
    \enddata
    \label{tab:forward_model_params}
    \tablecomments{Parameter definitions: metallicity is given in units of Solar metallicity. C/O ratio is given in absolute units. Gray cloud factor is in units of H$_2$ Rayleigh scattering cross-section at 350 nm. Haze factor is in units of nominal Rayleigh scattering. A heat redistribution value of 0.5 represents efficient redistribution and a value of 1.0 means no redistribution. $K_{zz}$ is the vertical turbulent diffusion parameter and $f_{\rm{sed}}$ is the sedimentation efficiency parameter. Units for logarithmic parameters refer to the argument.}
\end{deluxetable*}

\begin{deluxetable*}{lcccc}
    \renewcommand{\arraystretch}{1.1}
    \tabletypesize{\footnotesize}
    \tablecolumns{5} 
    \tablecaption{Summary of atmospheric retrievals.}
    \tablehead{ & \multicolumn{2}{c}{\textsc{POSEIDON}} & \multicolumn{2}{c}{\textsc{petitRADTRANS}}}
    \startdata
    Parameter \phantom{space} & \phantom{space}Prior\phantom{space} & \phantom{space}Posterior\phantom{space} & \phantom{space}Prior\phantom{space} &\phantom{space}Posterior\phantom{space} \\
    \hline
    $\log$\,g [cm/s$^2$] & $\cdots$ & $\cdots$ & $\mathcal{U}$(2.2, 3.0) &  $2.67^{+0.07}_{-0.07}$\\
    $T$ [K] & $\mathcal{U}$(400, 2300) & $1250^{+143}_{-113}$ & $\mathcal{U}$(1000, 2000) &  $1135^{+100}_{-85}$\\
    $R_{\rm{p, ref}}$ [R$_{\rm{J}}$] & $\mathcal{U}$(1.5895, 2.1505) & $1.72^{+0.004}_{-0.005}$ & $\mathcal{U}$(1.8, 2.1) &  $1.83^{+0.01}_{-0.01}$\\
    $\log$\,H$_2$O & $\mathcal{U}$(-12, -1) & $-4.53^{+0.37}_{-0.36}$ & $\mathcal{U}$(-13, -1) &  $-4.75^{+0.33}_{-0.29}$\\
    $\log$\,CH$_4$ & $\mathcal{U}$(-12, -1) & $-8.52^{+1.69}_{-2.23}$ & $\mathcal{U}$(-13, -1) &  $-10.21^{+2.30}_{-2.36}$\\
    $\log$\,CO$_2$ & $\mathcal{U}$(-12, -1) & $-5.40^{+0.56}_{-0.67}$ & $\mathcal{U}$(-13, -1) &  $-5.54^{+0.53}_{-0.72}$\\
    $\log$\,CO & $\mathcal{U}$(-12, -1) & $-7.61^{+2.53}_{-2.83}$ & $\mathcal{U}$(-13, -1) &  $-8.06^{+3.15}_{-3.84}$\\
    $\log$\,Na & $\mathcal{U}$(-12, -1) & $-9.20^{+1.91}_{-1.80}$ & $\mathcal{U}$(-13, -1) &  $-10.05^{+2.59}_{-2.49}$\\
    $\log$\,K & $\mathcal{U}$(-12, -1) & $-8.31^{+0.80}_{-1.74}$ & $\mathcal{U}$(-13, -1) &  $-10.02^{+1.87}_{-2.65}$\\
    $\log$\,SO$_2$ & $\cdots$ & $\cdots$ & $\mathcal{U}$(-13, -1) &  $-10.825^{+2.22}_{-2.17}$\\
    $\log$\,SiO$_2$ & $\mathcal{U}$(-30, -1) & $-12.55^{+0.75}_{-0.69}$ & $\cdots$ & $\cdots$ \\
    $\log$\,$r_{\rm{m}}$\,SiO$_2$ [\textmu m] & $\mathcal{U}$(-3, -1) & $-1.91^{+0.14}_{-0.16}$ & $\cdots$ & $\cdots$ \\
    $\log$\,$P_{\rm{cloud}}$\,SiO$_2$ [bars] & $\mathcal{U}$(-7, 2) & $-5.12^{+1.17}_{-1.19}$ & $\cdots$ & $\cdots$ \\
    $\Delta\log$\,$P$\,SiO$_2$ [bars] & $\mathcal{U}$(0, 9) & $2.36^{+1.65}_{-1.36}$ & $\cdots$ & $\cdots$ \\
    $\sigma_{\rm{lnorm}}$ [cm] & $\cdots$ & $\cdots$ & $\mathcal{U}$(1.05, 3.0) & $2.15^{+0.57}_{-0.66}$ \\
    $\log$\,$K_{zz}$ [cm$^2$/s] & $\cdots$ & $\cdots$ & $\mathcal{U}$(5.0, 13.0) & $6.36^{+1.26}_{-0.82}$ \\
    $f_{\rm{sed}}$ & $\cdots$ & $\cdots$ & $\mathcal{U}$(0.1, 10.1) & $4.74^{+3.24}_{-2.98}$ \\
    $\log$\,SiO$_{2\,\rm{base}}$ & $\cdots$ & $\cdots$ & $\mathcal{U}$(-10, 0) & $-5.75^{+0.77}_{-0.72}$ \\
    $\log$\,$P_{\rm{base}}$\,SiO$_2$ [bars] & $\cdots$ & $\cdots$ & $\mathcal{U}$(-10, 0) & $-3.10^{+0.63}_{-1.11}$ \\
    $\delta_{\rm{rel}}$ [ppm] & $\mathcal{U}$(-1000, 1000) & $122^{+61}_{-62}$ & $\mathcal{U}$(-52000, 52000) &  $118^{+71}_{-73}$ \\
    \enddata
    \label{tab:retrieval_model_params}
    \tablecomments{Parameter definitions: gravity, $g$; isothermal atmospheric temperature, $T$; reference radius, $R_{\rm{p, ref}}$; gas-phase volume mixing ratios, $\log$\,X; constant aerosol mixing ratio, $\log$\,SiO$_2$; mean particle size, $\log$\,$r_{\rm{m}}$\,SiO$_2$; cloud-top pressure, $\log$\,$P_{\rm{cloud}}$\,SiO$_2$; width of the cloud, $\Delta\log$\,$P$\,SiO$_2$; width of the particle-size distribution, $\sigma_{\rm{lnorm}}$; vertical mixing, $\log$\,$K_{zz}$; sedimentation efficiency, $f_{\rm{sed}}$; aerosol mixing ratio at the base, $\log$\,SiO$_{2\,\rm{base}}$; cloud-base pressure, $\log$\,$P_{\rm{base}}$\,SiO$_2$; MIRI dataset offset, $\delta_{\rm{rel}}$. Units for logarithmic parameters refer to the argument.}
\end{deluxetable*}

\begin{figure*}
\includegraphics[width=0.99\textwidth]{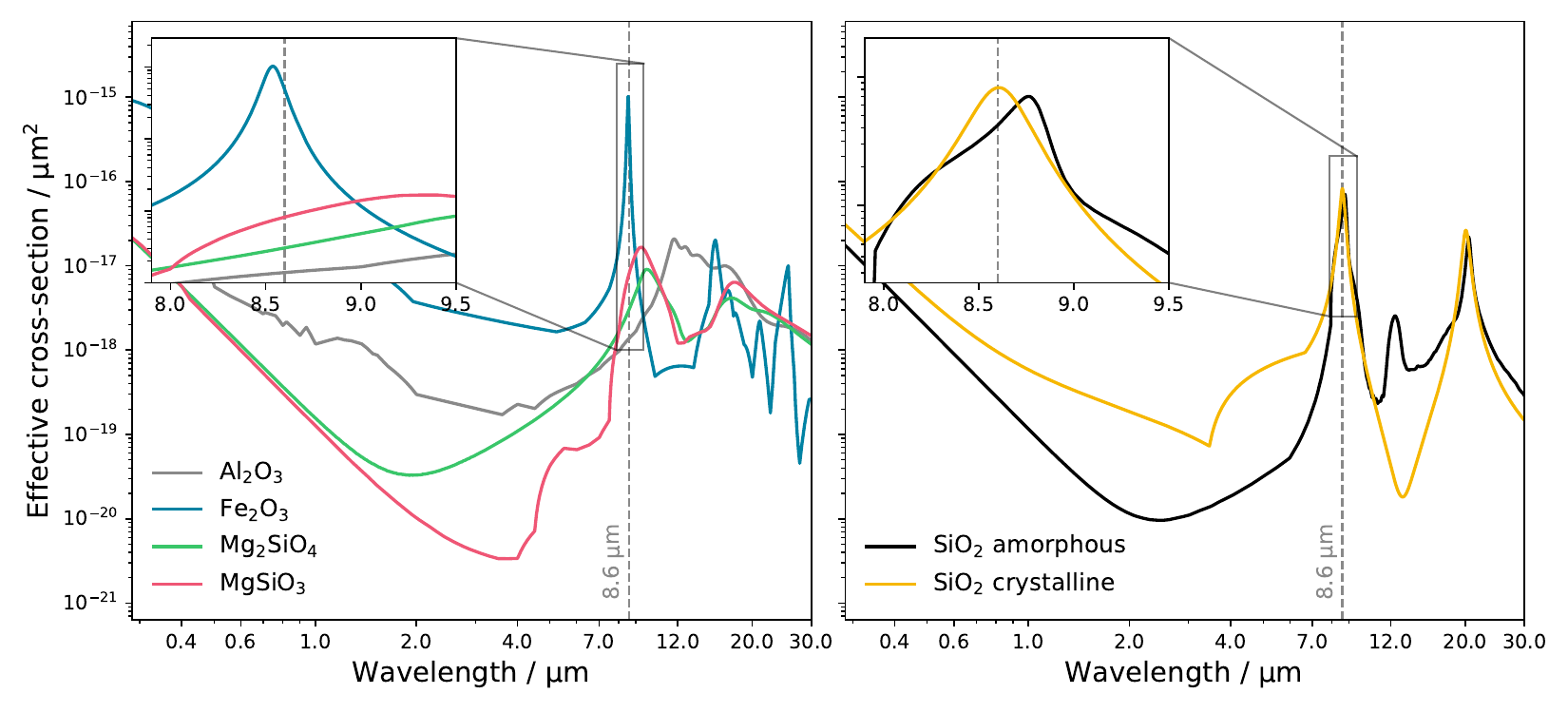}
\caption{Cloud condensate effective cross-sections. Left: candidate cloud species that were ruled out due to mismatching wavelengths of vibrational modes or unphysical particle sizes. Right: identified SiO$_2$(s) cloud species's crystalline versus amorphous forms. The inset panel shows how the crystalline form provides a better match to the 8.6\,\textmu m feature, whose wavelength is indicated by the dashed gray line.}
\label{fig:cross_sections}
\end{figure*}

\begin{figure*}
\includegraphics[width=0.99\textwidth]{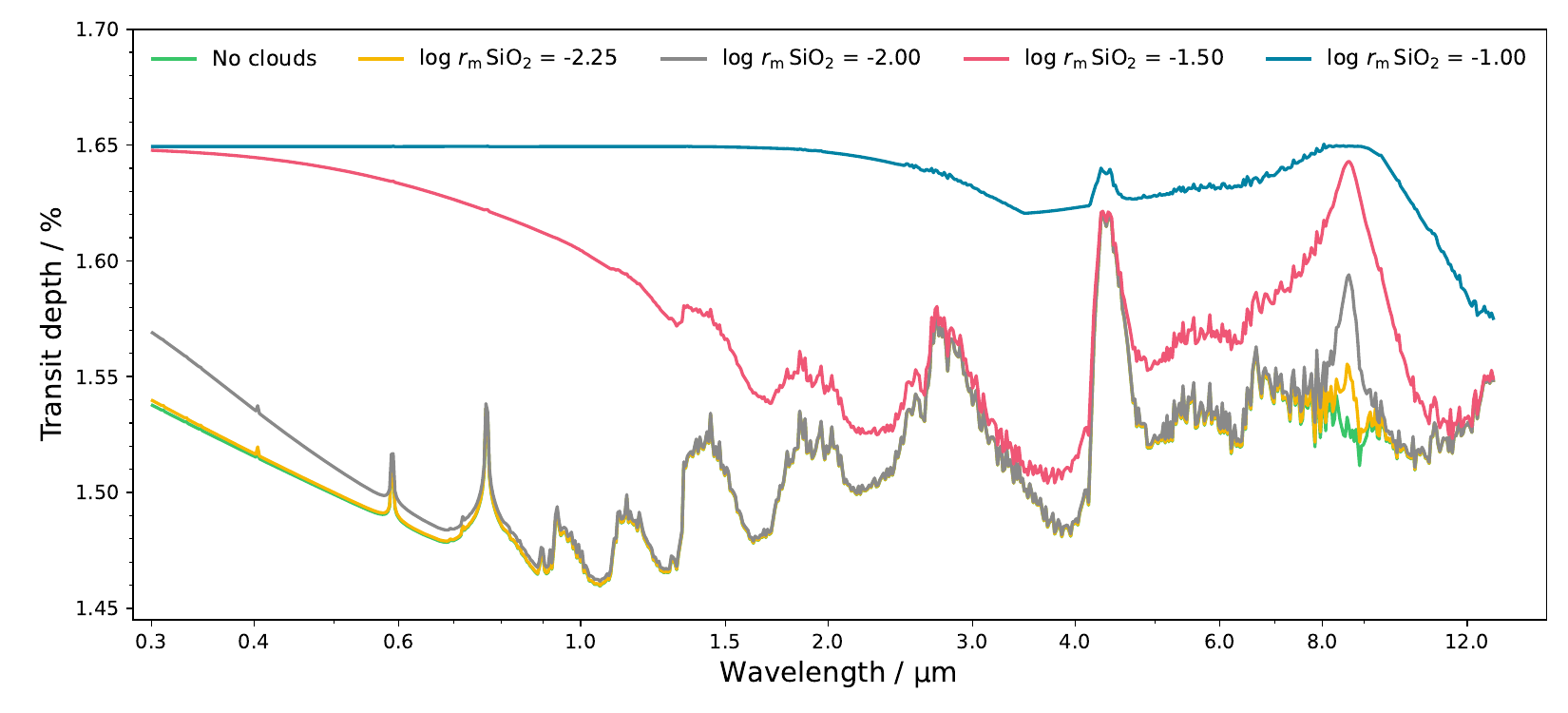}
\caption{Transmission spectra generated from the best-fit POSEIDON model, but for a range of different SiO$_2$(s) cloud particle sizes. The particle size directly affects both the scattering slope in the optical and the amplitude of the vibrational-mode absorption in the mid-infrared. The retrieved particle-size posterior is $\log$\,$r_{\rm{m}}$\,SiO$_2$ = $-1.91^{+0.14}_{-0.16}$.}
\label{fig:particle_sizes}
\end{figure*}

\bibliography{WASP17_JWST}{}
\bibliographystyle{aasjournal}

\end{document}